\documentclass[conference]{IEEEtran}
\usepackage{cite, xurl, graphicx, spverbatim}
\usepackage[hidelinks]{hyperref}
\usepackage[T1]{fontenc}

\begin{document}
\title{SBOM Challenges for Developers: \\ From Analysis of Stack Overflow Questions}

\author{
  \IEEEauthorblockN{
    Wataru~Otoda\IEEEauthorrefmark{1},
    Tetsuya~Kanda\IEEEauthorrefmark{2},
    Yuki~Manabe\IEEEauthorrefmark{3},
    Katsuro~Inoue\IEEEauthorrefmark{4},
    Yoshiki~Higo\IEEEauthorrefmark{1}
  }
  \IEEEauthorblockA{
    \IEEEauthorrefmark{1}
    Osaka University, Japan,
    \{wa-otoda, higo\}@ist.osaka-u.ac.jp
  }
  \IEEEauthorblockA{
    \IEEEauthorrefmark{2}
    Notre Dame Seishin University, Japan,
    kanda@m.ndsu.ac.jp
  }
  \IEEEauthorblockA{
    \IEEEauthorrefmark{3}
    The University of Fukuchiyama, Japan,
    manabe-yuki@fukuchiyama.ac.jp
  }
  \IEEEauthorblockA{
    \IEEEauthorrefmark{4}
    Nanzan University, Japan,
    inoue599@nanzan-u.ac.jp
  }
}

\maketitle

\begin{abstract}
  Current software development takes advantage of many external libraries, but it entails security and copyright risks. While the use of the Software Bill of Materials (SBOM) has been encouraged to cope with this problem, its adoption is still insufficient. In this research, we analyzed the challenges that developers faced in practicing SBOM use by examining questions about SBOM utilization on Stack Overflow, a Q\&A site for developers. As a result, we found that (1) the proportion of resolved questions about SBOM use is 15.0\% which is extremely low, (2) the number of new questions has increased steadily from 2020 to 2023, and (3) SBOM users have three major challenges on SBOM tools.
\end{abstract}


\section{Introduction}
Modern software development is often accomplished not by implementing all functionalities from scratch but by leveraging numerous external libraries. This approach not only reduces development costs and shortens development cycles but also allows for the use of mature software components, leading to robust and sophisticated software \cite{study-software-reuse,managing-software-reuse}. However, such activity comes with risks concerning cyber security and copyright \cite{library-usages-updates-risks,oss-supply-chain-attacks,oss-license-violation}. Indeed, in 2020, a cyber incident occurred where malware was embedded in the software update of SolarWinds' Orion software through a supply chain attack, leading to cyberattacks on numerous U.S. government agencies and private sectors that used Orion software \cite{usgov-solarwinds}. Despite identifying and addressing these risks early on is crucial, the immense number of external libraries used in current software development, especially considering transitive dependencies like dependencies of dependencies \cite{dependency-network}, makes proper management challenging.

To address this issue, the use of the Software Bill of Materials (SBOM) is encouraged. SBOM is a formal and machine-readable record of software components (e.g. libraries) used in building software, which includes details such as licenses, versions, and vendors of each software component, as well as information on supply chain relationships between components \cite{ntia-sbom-min-elems,linux-foundation-readiness}. CycloneDX and SPDX are the primary formats for representing SBOM \cite{sbom-history}. While SBOM is rapidly getting attention in recent years with government agencies promoting its use \cite{us-exec-order,eu-cyber-resilience-act}, its adoption is still insufficient \cite{sbom-study,sbom-study-github}. While there are studies identifying causes and proposing solutions through surveys conducted on companies and organizations, no study has analyzed the challenges faced by ordinal developers in utilizing SBOM in practice. Therefore, this study analyzes questions on the use of SBOM on Stack Overflow, a primary question-and-answer (Q\&A) site for developers, where we deemed developers facing challenges in utilizing SBOM likely to seek solutions on by asking questions, considering its popularity \cite{so-popularity}.

The remainder of this paper is organized as follows. Section \ref{sec:related} describes related works. Section \ref{sec:study} explains the research methodology. Section \ref{sec:result} reports the results and discussion. Finally, section \ref{sec:conclusion} concludes the research and discusses future works.

\section{Related Work}
\label{sec:related}

The Linux Foundation conducted an SBOM readiness survey in 2022 targeting 412 organizations of various industry types and sizes \cite{linux-foundation-readiness}. While 76\% of organizations work on SBOM adoption, the survey revealed concerns such as the industry's honest commitment to SBOM adoption, the lack of industry consensus on what an SBOM should contain, and the value to their customers by providing them with an SBOM being unclear. Xia et al. conducted an interview on the adoption status and challenges of SBOM among developers, highlighting the importance of incentives for SBOM generation, industry consensus on what to include in SBOM, mechanisms for selective sharing of SBOM content, SBOM content validation/verification, mature SBOM tools, and increasing awareness of SBOM \cite{sbom-study}. Stalnaker et al. investigated the challenges stakeholders face during SBOM creation and usage, emphasizing the importance of multi-dimensional SBOM specifications, enhanced SBOM tooling and build system support, SBOM content validation, and incentives for SBOM adoption \cite{bomsaway-stakeholders-study}.

While these studies provide insights into the current status and future challenges of SBOM adoption in the industry, they do not target ordinal software developers outside of companies or organizations. In contrast, Nocera et al. analyzed the SBOM adoption on open-source software (OSS) repositories on GitHub, the most popular version control platform for both personal and professional use \cite{sbom-study-github}. The results indicated a low adoption of SBOM in OSS repositories, with few repositories meeting the guidelines set by the National Telecommunications and Information Administration (NTIA) for how SBOM should be supplied, highlighting the importance of convenient SBOM generation tools that can be easily integrated into Continuous Integration and Continuous Delivery (CI/CD) pipelines and build-automation tools. However, this study provides quantitative insights and does not reveal the concrete challenges software developers face.

\section{Analysis}
\label{sec:study}

\begin{figure}[!t]
  \begin{minipage}{\linewidth}
    \centering
    \includegraphics[width=0.7\linewidth]{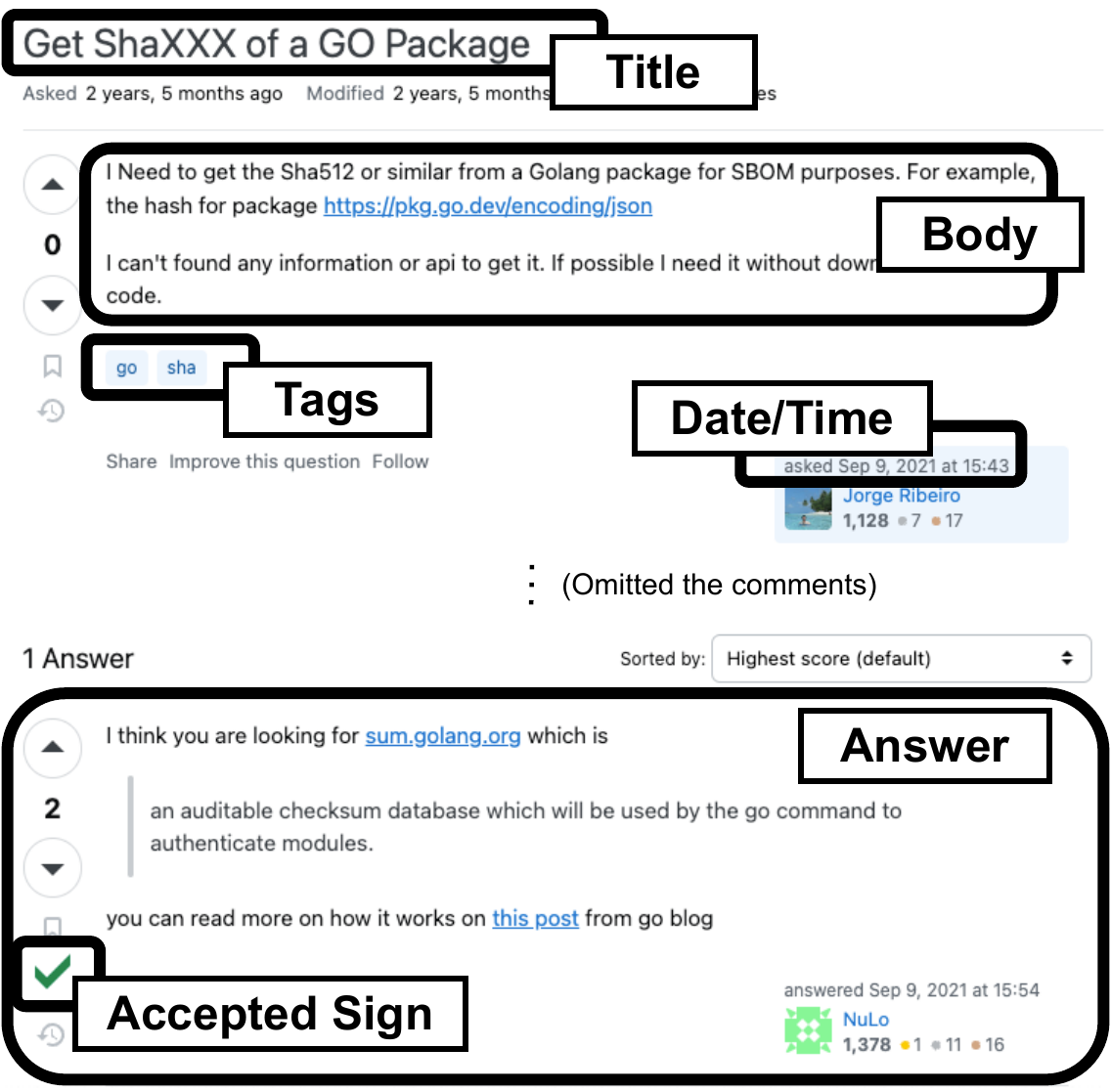}
    \caption{A question found on Stack Overflow.}
    \label{fig:so-image}
  \end{minipage}
\end{figure}

In this study, we analyzed questions related to SBOM uses on Stack Overflow, focusing on the CycloneDX and SPDX formats. Stack Overflow is a developer Q\&A site where users can post and search for questions and answers. \figurename~\ref{fig:so-image} provides an example of a question on Stack Overflow\footnote{\url{https://stackoverflow.com/questions/69121175}}. A post on Stack Overflow typically consists of a title, question, multiple tags, and multiple answers to the question. Among answers, the one deemed most helpful by the question asker in resolving the issue is marked with a green checkmark indicating the \emph{Accepted} status. However, no answers may get accepted in such a case as all of them fail to resolve the question.

\begin{figure}[!t]
  \centering
  \includegraphics[width=0.4\linewidth]{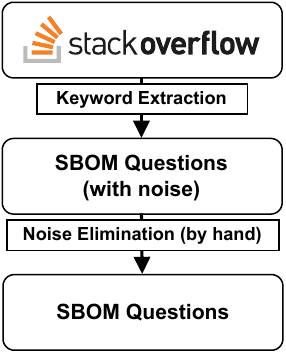}
  \caption{Flow of this analysis.}
  \label{fig:study-flow}
\end{figure}

We followed the process outlined in \figurename~\ref{fig:study-flow}. First, we selected search keywords that are \verb|SBOM|, \verb|CycloneDX|, and \verb|SPDX|. Then, we extracted questions that case-insensitively contained at least one of these keywords in the title or body. Although Stack Overflow allows users to set tags based on the technical domains of their questions, SBOM-related tags were not widely used at the time of the analysis. We then filtered out non-SBOM-related questions (referred to as noise) e.g. those containing a log text from an unrelated framework or program code that happens to contain an SBOM-related word in it.

For the keywords \verb|SBOM| and \verb|CycloneDX|, we first entered \verb|SBOM is:question| and \verb|CycloneDX is:question| into the search bar on the Stack Overflow site to obtain SBOM questions and then removed noise by hand. On the other hand, for the keyword \verb|SPDX|, we used the Stack Exchange Data Dump 2023-09-12\footnote{\url{https://archive.org/details/stackexchange_20230912}}. This was because the search results exceeded the limit of 500 questions imposed by Stack Overflow's search function. We filtered out posts that contained noise pattern strings at least once mechanically and then we removed the remaining noise by hand. Table. \ref{tab:noise-spdx} shows the noise pattern strings and their frequencies in the \verb|SPDX| keyword search.

\begin{table}[!t]
  \caption{Noise Pattern Strings for the Keyword \emph{SPDX}.}
  \label{tab:noise-spdx}
  \centering
   \begin{tabular}{c||c}
    \hline
    \textbf{Noise Pattern Strings} & \textbf{Count} \\
    \hline\hline
    \verb|SPDX-License-Identifier| & 1229 \\
    \hline
    \begin{tabular}{c}
      \verb|License should be a valid| \\
      \verb|SPDX license expression|
    \end{tabular} & 30 \\
    \hline
    \verb|spdy| & 51 \\
    \hline
    \begin{tabular}{c}
      \verb|spdx-correct|, \\
      \verb|spdx-expression-parse|, \\
      \verb|spdx-exceptions|, \\
      \verb|spdx-license-ids|
    \end{tabular} & 62 \\
    \hline
  \end{tabular}
\end{table}

The noise pattern strings were selected from frequent texts in noise and verified to not appear in an SPDX-formatted SBOM. The reasons why we deemed them noise pattern strings are as follows. \verb|SPDX-License-Identifier| is a standardized notation used to declare licenses at the beginning of program code, which is specified as part of the SPDX standard \cite{spdx-identifier-in-source} but the notation itself is not an SBOM. Similarly, \spverb|License should be a valid SPDX license expression| is a warning message output by the Yarn package manager when a package's license notation is incorrect. \verb|spdy| was selected because it appeared in cases where \verb|spdx| (with varying capitalization) was used as a variable name representing the velocity on the $x$-axis in programs, often accompanied by \verb|spdy|, representing the velocity on the $y$-axis. \verb|spdx-correct|, \verb|spdx-expression-parse|, \verb|spdx-exceptions|, and \verb|spdx-license-ids| are package names on the NPM package manager, typically appearing in log outputs contained in questions regarding NPM errors where these packages are included in the project dependencies.

As a result of this noise elimination process, 27 questions out of 58 remained for the keyword \verb|SBOM|, 28 out of 31 for \verb|CycloneDX|, and 6 out of 1432 for \verb|SPDX|, totaling 42 questions due to duplicates. From these questions, we performed the following analyses:

\subsubsection*{Analysis 1}
We examined the presence of answers and the resolution status of each question. Questions with an accepted answer were considered resolved, while those without were considered unresolved.

\subsubsection*{Analysis 2}
We summed up the number of new questions posted each year.

\subsubsection*{Analysis 3}
We extracted technical issues from the content of each question, identifying frequently asked questions and those seemingly unresolved based on the subjective judgment of the lead author.

\section{Analysis Results and Discussion}
\label{sec:result}

\subsection{Analysis 1: Answered and Resolved Rate of SBOM Questions}
\begin{figure}[!t]
  \centering
  \includegraphics[width=\linewidth]{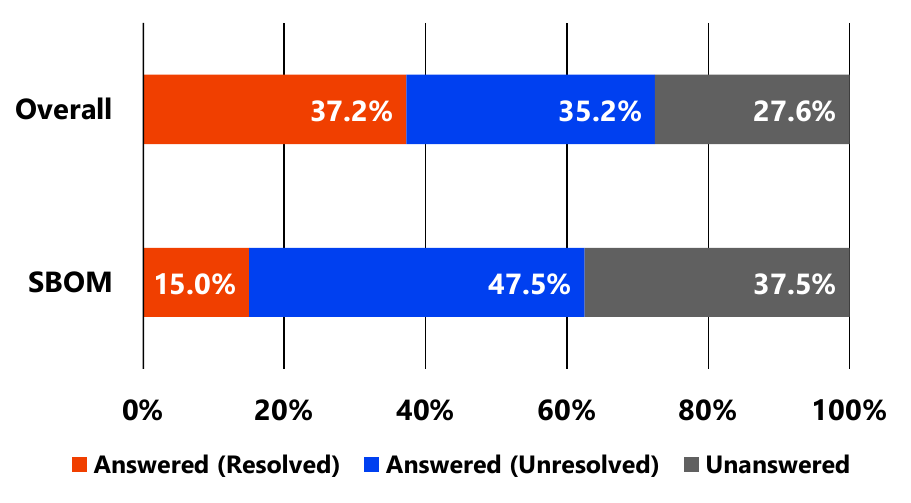}
  \caption{Answered and Resolved Rate of SBOM Questions.}
  \label{fig:q-resolved-result}
\end{figure}

\figurename~\ref{fig:q-resolved-result} shows the proportion of \emph{Answered (resolved)}, \emph{Answered (unresolved),} and \emph{Unanswered} SBOM questions, along with that of overall questions on Stack Overflow. To facilitate a fair comparison, we filtered out questions posted before 2021 when SBOM questions were not frequently asked. Among the total of 40 SBOM questions, 6 questions (15.0\%) were answered (resolved), 19 questions (47.5\%) were answered (unresolved), and 15 questions (37.5\%) were unanswered.

Considering that the proportion of unanswered questions overall is 27.6\%, SBOM questions are answered likely on average. However, the resolution rate of SBOM questions is at an extremely low level from the proportion of resolved questions in the overall being 37.2\%. It suggests that it is challenging to find solutions on Stack Overflow when SBOM users face issues, indicating either a shortage of software developers knowledgeable enough to answer SBOM questions or SBOM tools being so immature that many SBOM issues are currently unsolvable at the time of questioning, as also discussed in Section \ref{sec:study-bugs}.

\subsection{Analysis 2: Trends in the Number of SBOM Questions}
\begin{figure}[!t]
  \centering
  \includegraphics[width=\linewidth]{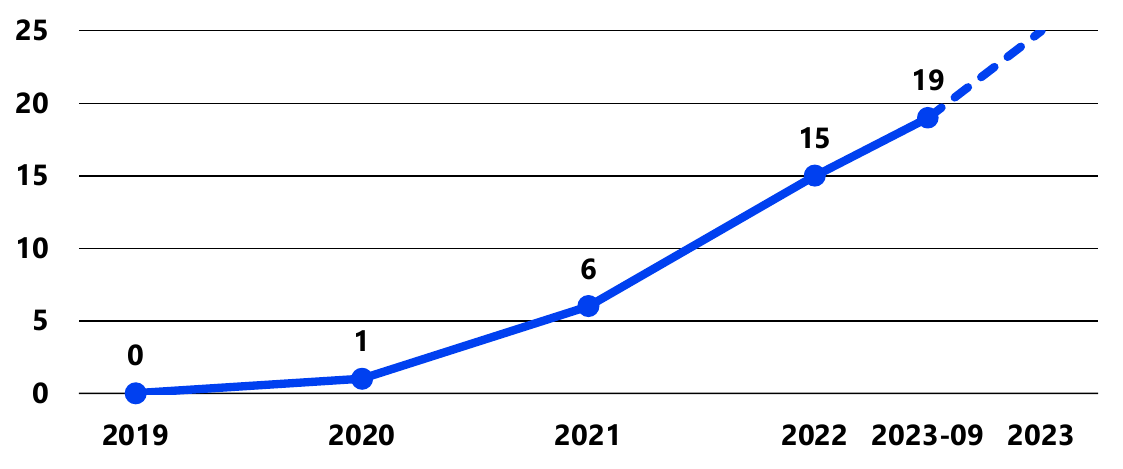}
  \caption{Annual Trends in the Number of New SBOM Questions.}
  \label{fig:num-questions-result}
\end{figure}

The annual trends in the number of new SBOM questions from 2019 to 2023 as of September 2023 are shown in \figurename~\ref{fig:num-questions-result}. Before 2019, the number of SBOM questions was 1 in 2012 and 0 for the rest.

Considering the popularity of Stack Overflow, SBOM questions extracted in this analysis remained small over the four years. However, we can observe a steady increase in the number of new SBOM questions from 2020 to 2023, with a notable acceleration from 2021 onwards, possibly correlated with the issuance of the U.S. executive order requesting SBOM use and the standardization of SPDX by ISO/IEC in 2021.

\subsection{Analysis 3: Challenges Faced by SBOM Users}
We found that the following challenges are significant:

\begin{enumerate}
  \item Insufficient coverage of use cases by SBOM tools (7 posts)  
  \item Inability of SBOM tools to meet requirements (4 posts)  
  \item Immaturity of SBOM tools or unclear usages (19 posts)  
\end{enumerate}

Below are actual questions found on Stack Overflow highlighting each challenge.

\subsubsection{Insufficient Coverage of Use Cases by SBOM Tools}
There is a question titled \emph{``How do I generate a Cyclonedx bom for a Java project built with Ant?''} posted on March 2022\footnote{\url{https://stackoverflow.com/questions/71605182}}, which is from a software developer with experience generating an SBOM for Gradle projects but facing difficulties with an older Apache Ant project. While an answer suggests that CycloneDX Core (Java) might work, it remains unaccepted. This question highlights insufficient tool support for older project management systems. Another question titled \emph{``Is there any tool through which we can generate SBOM report (SPDX / CycloneDX) for Windows programs?''} on September 2022\footnote{\url{https://stackoverflow.com/questions/73648096}} also highlights the absence of tools for generating SBOMs for applications installed on Microsoft Windows, despite the availability of such tools for Linux. While answers exist, none have been accepted. The lack of such tools could be attributed to the decentralized nature of software metadata management on Windows compared to Linux, making it challenging to develop SBOM generation tools for Windows.

\subsubsection{Inability of SBOM tools to Meet Requirements}
There is a question titled \emph{``NTIA minimum SBOM requirement tool''} posted on April 2023\footnote{\url{https://stackoverflow.com/questions/76103711}}, which illustrates a developer's attempt at generating SBOMs for Java, Python, iOS (Swift), and Android (Kotlin) projects while adhering to the NTIA's guideline. Despite trying various tools, none fulfilled all the required fields, indicating a gap between SBOM tools' ability and compliance requirements or challenges in obtaining all the necessary information as specified in the guideline from the perspective of SBOM tool developers. Another question titled \emph{``How to include Open Source license in GitHub SBOM export?''} on August 2023\footnote{\url{https://stackoverflow.com/questions/76962834}} also addresses the absence of open source license information in GitHub's SBOM export, highlighting the limitations of GitHub's SBOM generation functionality for compliance management purposes.

\subsubsection{Immaturity of SBOM Tools or Unclear Usages}
\label{sec:study-bugs}
There is a question titled \emph{``How to create a BOM file in a Flutter project''} posted on June 2023\footnote{\url{https://stackoverflow.com/questions/76515339}}, which details a developer's struggle to generate a CycloneDX SBOM for an Android portion of a Flutter project, encountering errors during the process. Similar questions addressing errors and unclear usages by SBOM generation tool users were prevalent. These questions indicate that SBOM generation tools are relatively immature, leading to defects, insufficient features, and a lack of comprehensive documentation or accumulated knowledge among software developers on basic usage, making the correct SBOM generation difficult.

Through this analysis, questions requiring fundamentally new approaches or systems were scarce, with many revolving around usages, issues, and minor missing functionalities on existing tools. This suggests that developers seek accumulated knowledge, refinement, and extension to existing tools.

\subsection{Threats to Validity}
\label{sec:validity}

We filtered out questions using the noise pattern strings listed in Table. \ref{tab:noise-spdx} when analyzing questions containing the keyword \verb|SPDX|. However, by doing so, there is a possibility that questions related to SBOM use, which should not have been excluded as noise, were excluded. However, in selecting noise pattern strings, we confirmed that posts containing those strings were likely noise by extracting multiple posts containing the string for investigation. Therefore, the threat to validity caused by this exclusion is considered small.

On another note, the number of questions extracted as meeting the criteria was only 42. While this is possibly due to the insufficient popularity of SBOM, it might affect the accuracy of the analysis results. Although there are also general Q\&A sites like Quora apart from Stack Overflow, we observed no active usage of these platforms by developers or managers involved in SBOM adoption.

\section{Conclusion and Future Work}
\label{sec:conclusion}

In this research, we analyzed questions posted on the developer Q\&A site Stack Overflow on the use of SBOM. Through analyzing the answered and resolved rate of SBOM questions (Analysis 1), we revealed that Stack Overflow is still not a satisfactory venue to ask SBOM questions to obtain solutions. The second analysis on trends in the number of SBOM questions (Analysis 2) revealed a notable increase from 2021, possibly due to the U.S. executive order mandating SBOM use and the standardization of SPDX by ISO/IEC. The last analysis on challenges faced by SBOM users (Analysis 3) identified three main challenges faced by SBOM users.

We consider it a future work to keep tracking the latest trends in SBOM use, in such a way as to continue investigating Stack Overflow which is showing an increasing trend in the number of SBOM questions, or to conduct analyses on websites other than Q\&A sites such as Reddit or issues in the SBOM-related projects on GitHub.

\section*{Acknowledgment}
This work was supported by JSPS KAKENHI Grant Numbers
JP24K14895, 
JP21K02862, 
JP23K28065, 
JP21K18302, JP21H04877, JP22H03567, JP22K11985, 
and Nanzan University Pache Research Subsidy I-A-2 for the 2024 academic year. 

\bibliographystyle{IEEEtran}
\bibliography{IEEEabrv, refs}
\end{document}